# Imaging Intravoxel Vessel Size Distribution in the Brain Using Susceptibility Contrast Enhanced MRI

Natenael B. Semmineh,[1*] Indranil Guha,[1*] Deborah Healey,[1]
Anagha Chandrasekharan,[1] Jerrold L. Boxerman,[4] C. Chad Quarles[1,2,3]

[1]Department of Cancer Systems Imaging, Cancer Neuroscience Program, Cancer Neuroimaging Research Program,

[2]Department of Neuroradiology,

[3]Department of Neurooncology, The University of Texas MD Anderson Cancer Center, Houston, TX, 77025, USA

[4]Department of Diagnostic Imaging, Rhode Island Hospital, Providence, RI, 02903, USA

**Corresponding Author:** C. Chad Quarles

**Address:** Cancer Systems Imaging, 1881 East Road, Unit 1907, Houston, TX 77054-1907, USA

**Phone:** (+1) 832-885-8540

**Email:** ccquarles@mdanderson.org

* These authors share first authorship

**ABSTRACT**

Vascular remodelling is inherent to the pathogenesis of many diseases including cancer, neurodegeneration, fibrosis, hypertension, and diabetes. In this paper, a new susceptibility-contrast based MRI approach is established to non-invasively image intravoxel vessel size distribution (VSD), enabling a more comprehensive and quantitative assessment of vascular remodelling. The approach utilizes high-resolution light-sheet fluorescence microscopy (LSFM) images of rodent brain vasculature, gradient echo sampling of free induction decay and spin echo (GESFIDE) MRI signal simulation from the three-dimensional (3D) vascular networks, and training a deep learning (DL) model to predict cerebral blood volume (CBV) and VSD from GESFIDE signals. Specifically, small voxel-size volumes of interest (VOI) (n=32,000) were extracted from LSFM images of rodent brain and the vascular structure was segmented. Next, two DL models were trained to predict the CBV and VSD from the ratio of pre- and post-contrast GESFIDE signals simulated from these VOIs. The results from *ex vivo* experiments on test VOIs (n=3,132) demonstrated strong linear correlation (r=0.95) and high similarity (mean Bhattacharya Coefficient (BC)=0.87) between the true and predicted CBV and VSDs, respectively. The DL models outperformed the traditional dictionary-matching approach and demonstrated high accuracy in predicting CBV and VSD, even when the GESFIDE signals were degraded with varying noise levels (SNR: 15, 30, 45, and 60 dB). The DL model showed comparable results to those observed in the test VOIs on a public mouse brain vasculature dataset (n=1,000), demonstrating the generalizability of the DL models. The accuracy of the predicted CBV (r=0.78) and VSD (mean BC = 0.82) on the tumor VOIs (n=706) were moderately high but lower than the accuracy of predicted CBV and VSD observed for the healthy VOIs. Hence, with further *in vivo* validation, intravoxel VSD imaging could become a transformative preclinical and clinical tool for interrogating disease and treatment induced vascular remodelling.

**Keywords:** Vessel fingerprinting, GESFIDE signal, deep learning, cerebral blood volume, vessel size distribution.

## 1. INTRODUCTION

Vessel size distribution (VSD) is a fundamental feature of vascular architecture, tightly linked to physiological function, metabolism, and pathological processes. In healthy tissues, vascular organization is finely regulated to meet organ-specific metabolic demands, ensuring efficient blood flow and oxygen delivery. In disease, this balance is disrupted and aberrations in vascular architecture become defining features of many pathological conditions. For example, tumor angiogenesis leads to disordered vascular morphology and networks (Carmeliet & Jain, 2000; Gatenby & Gillies, 2008; Hanahan & Weinberg, 2011; Nagy et al., 2012), while luminal narrowing and capillary rarefaction restrict blood flow in atherosclerosis (Ross, 1999). Diabetic microangiopathy is characterized by injury to arterioles and venules along with pro- and anti-neovascularization, leading to premature blood vessels and micro-thrombosis (Bataller & Brenner, 2005). Cerebral small-vessel disease, a major contributor to stroke and dementia is marked by capillary rarefaction, vessel narrowing, and autoregulatory dysfunction, leading to chronic hypoxia, impaired cerebral blood flow, and cognitive decline (Østergaard et al., 2016; Pantoni, 2010). Microvascular damage in Alzheimer's disease impairs cerebral blood flow, promotes amyloid-beta accumulation, and contributes to cognitive decline (Iadecola, 2004; Zlokovic, 2011), whereas in Parkinson's disease, vessel remodelling and reduced blood flow lead to neuronal damage (Paul & Elabi, 2022; Sweeney et al., 2018). Given the widespread impact of vascular abnormalities across multiple organ systems, the non-invasive determination of intravoxel vessel size heterogeneity could be a transformative tool for tissue and disease characterization, mechanistic explorations, diagnostics, and treatment response assessment in both animals and humans.

Accurately and non-invasively imaging VSD remains an unmet challenge, with current efforts relying on *ex vivo* microscopy, which though informative, is limited by sampling constraints and unsuitability for *in vivo* or longitudinal studies. MRI enables non-invasive and multi-organ imaging, while contrast agent (CA) enhancement further allows for the assessment of tissue structure and function through pre- and post-injection MR imaging. After CA administration, the decrease in longitudinal (T1) and transverse (T2 and $T2^*$) relaxation times of tissue water is, in part, determined by the CA concentration. Regarding CA-induced T2 and $T2^*$ changes, when CA is introduced into blood vessels, it creates a susceptibility difference between the vessels and the surrounding tissue. The susceptibility difference generates magnetic field inhomogeneities surrounding the blood vessels (and whose magnitude depends on the vascular architecture) leading to enhanced proton dephasing in the extravascular space and a decrease in the transverse relaxation times.

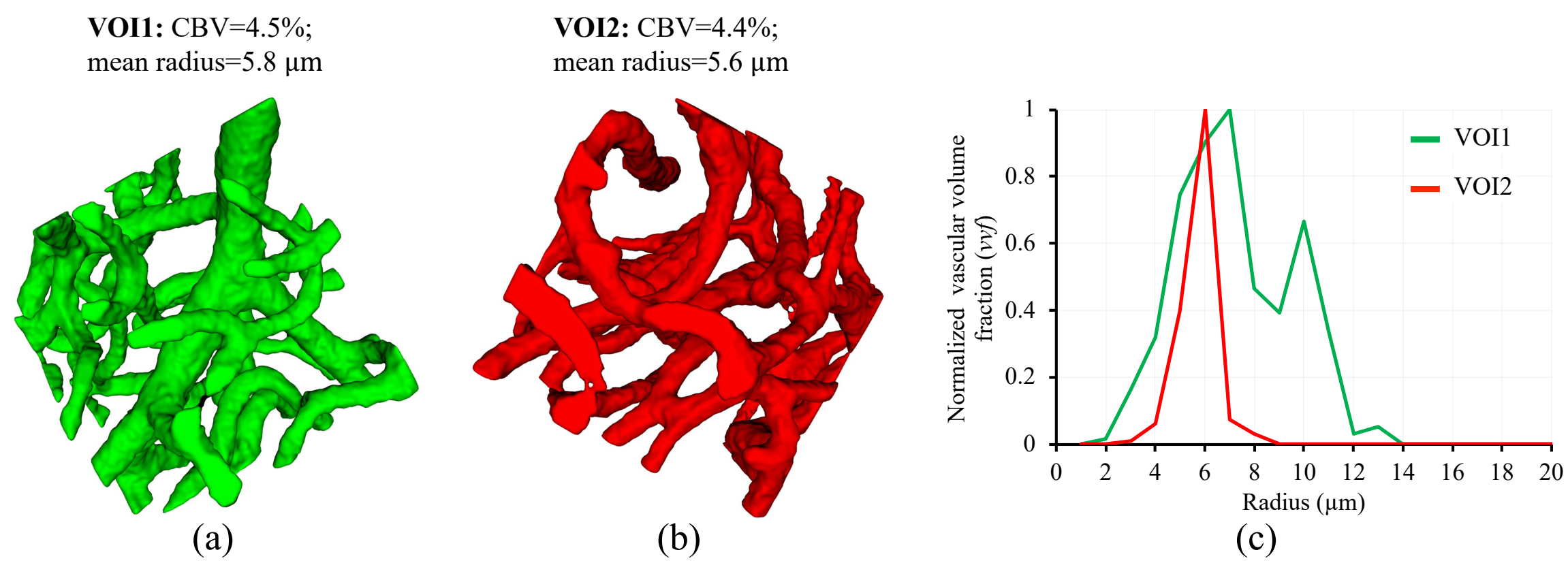

Figure 1. Two representative VOIs (a,b) extracted from LSFM image of a mouse-brain vasculature with similar CBV and mean radius but different VSDs (c). Specifically, VOI1 showed a broader VSD, indicating greater variability in vessel sizes compared to VOI2.

The most common technique relying on $T_2$ and $T_2^*$ changes, dynamic susceptibility contrast (DSC)-MRI, employs gradient-echo (GE) or spin-echo (SE) acquisitions to measure changes in transverse relaxation rates ($\Delta R_2^*$ and $\Delta R_2$), enabling the computation of perfusion parameters (Østergaard et al., 1996). Simulations and experimental data have shown that the change in the GE relaxation rate ($\Delta R_2^*$) initially increases for very small perturber sizes and then plateaus as perturber size increases. In contrast, the SE relaxation rate change ($\Delta R_2$) increases, peaks, and then decreases with a maximal sensitivity towards capillary sized perturbers (Boxerman et al., 1995). Analytical methods were first proposed to measure the MRI signal decay only due to static dephasing (Yablonskiy & Haacke, 1994) which was later extended to include diffusion effects and signal contribution from intravascular spins (Kiselev & Posse, 1999). These models laid the foundation for modelling the signal relaxation time course of both the free induction decay (FID) and the SE experiments. When a simultaneous GE and SE sequence is utilized, measures of mean vessel size within a voxel can be derived, an approach termed as vessel size imaging (Dennie et al., 1998). Biophysically, vessel size imaging relies upon the differential vessel size sensitivity of GE and SE signal. The ratio of $\Delta R_2^*$ and $\Delta R_2$ was first used as a relative measure of mean vessel diameter (Dennie et al., 1998), while the Q-index (Q=$\Delta R_2/\Delta R_2^{*\frac{2}{3}}$) was later introduced as a measure of microvascular density (Jensen & Chandra, 2000). Analytical models for computing vessel size index (VSI) and mean vessel radius as function of the ratio of $\Delta R_2^*$ and $\Delta R_2$, the apparent diffusion coefficient, and the susceptibility difference ($\Delta\chi$) have been developed and validated on animals (Tropres et al., 2001) and humans (Kiselev et al., 2005). However, these models can exhibit reduced accuracy due to high $\Delta\chi$ assumptions and simplified representations of vessel shape and water diffusion.

A more sophisticated approach for quantifying mean vessel size, blood volume, and oxygenation, termed MR vascular fingerprinting (MRvF), was proposed to overcome prior assumptions and enable higher resolution vessel size imaging (Boux et al., 2021; Christen et al., 2014; Delphin et al., 2024; Pouliot et al., 2017). Unlike MR fingerprinting (MRF), where signals are obtained using a pseudorandomized variation of acquisition parameters, such as flip angle and repetition time, to generate a unique fingerprint of signal evolution for each tissue type, MRvF utilizes a fixed MR sequence, specifically GE sampling of the free induction decay and spin echo (GESFIDE), and injection of an iron-oxide nanoparticle based contrast agent, to interrogate vascular morphological parameters, including cerebral blood volume (CBV), vessel radius, and oxygenation within a voxel. The first MRvF technique (Christen et al., 2014) relied upon a dictionary of simulated GESFIDE signal pre- and post-injection of an iron-based CA using virtual voxels containing two-dimensional (2D) blood vessels with varying CBV, mean vessel radius, and blood oxygenation saturation ($SO_2$). The dictionary was used to predict vascular parameters for any given MRI signal and the experimental results on the healthy human brain showed that the parametric maps predicted from the MRvF was consistent with the same maps obtained from the conventional MR methods. A computationally efficient version of MRvF (Boux et al., 2021) was later proposed that used novel dictionary-based statistical learning method to estimate vascular parameters from MRvF with higher accuracy. Realistic cortex angiograms of mouse have also been used instead of synthetic vessel models for MRvF (Pouliot et al., 2017) to better capture vascular complexity. It was found that parameter estimates were biased when different angiograms were used for dictionary matching, but the method improved physiological accuracy over 2D models. This approach revealed significantly lower $SO_2$, CBV, and mean vessel radius in atherosclerotic mice compared to the wild type mice. Nevertheless, simple cylindrical vascular models insufficiently capture microvascular characteristics in pathological conditions, where vessel networks exhibit varying degrees of anisotropy and pronounced differences in vessel shape and tortuosity. Recently, the original MRvF technique was extended (Delphin et al., 2024) using three-dimensional (3D) vascular structures extracted from microscopic images of whole mouse brain vasculature and the mean radius, blood volume fraction, and $SO_2$ estimates obtained using this approach showed better agreement with the literature than same measures obtained using 2D or 3D cylindrical models. However, the paper did not report quantitative metrics demonstrating the agreement between the true and predicted parameters.

The advantage of GESFIDE lies in its sensitivity to a broader spectrum of microstructural variations by integrating GE, asymmetric spin echo (ASE), and SE contrasts. With respect to the vasculature, GE is sensitive to vessels of all sizes, SE is primarily sensitive to capillary sized vessels, and ASE provides an intermediate sensitivity. GESFIDE effectively combines these contributions into a more comprehensive vascular fingerprint. However, despite the enhanced sensitivity of GESFIDE, both VSI and MRvF provide mean vascular parameters within an MRI voxel that does not reflect the underlying (intravoxel) heterogeneity of the vascular architecture. For example, two volumes of interest (VOIs) (Figure 1(a,b)) extracted from light-sheet fluorescence microscopy (LSFM) images of a cleared whole mouse brain vasculature with similar mean radius (VOI1: 5.8 µm, VOI2: 5.6 µm) and CBV (VOI1: 4.5%, VOI2: 4.4%) exhibited distinct VSDs (1(c)). In this study, VSD was defined as the normalized vessel volume fraction (*vvf*) weighted histogram of vessel radius values with a bin size of 1µm. The VSD of VOI1 was broader, indicating a more heterogeneous distribution of vessel sizes compared to VOI2. This highlights that while mean vascular parameters provide a useful summary, VSD offers deeper insight into the underlying heterogeneity of the vascular network.

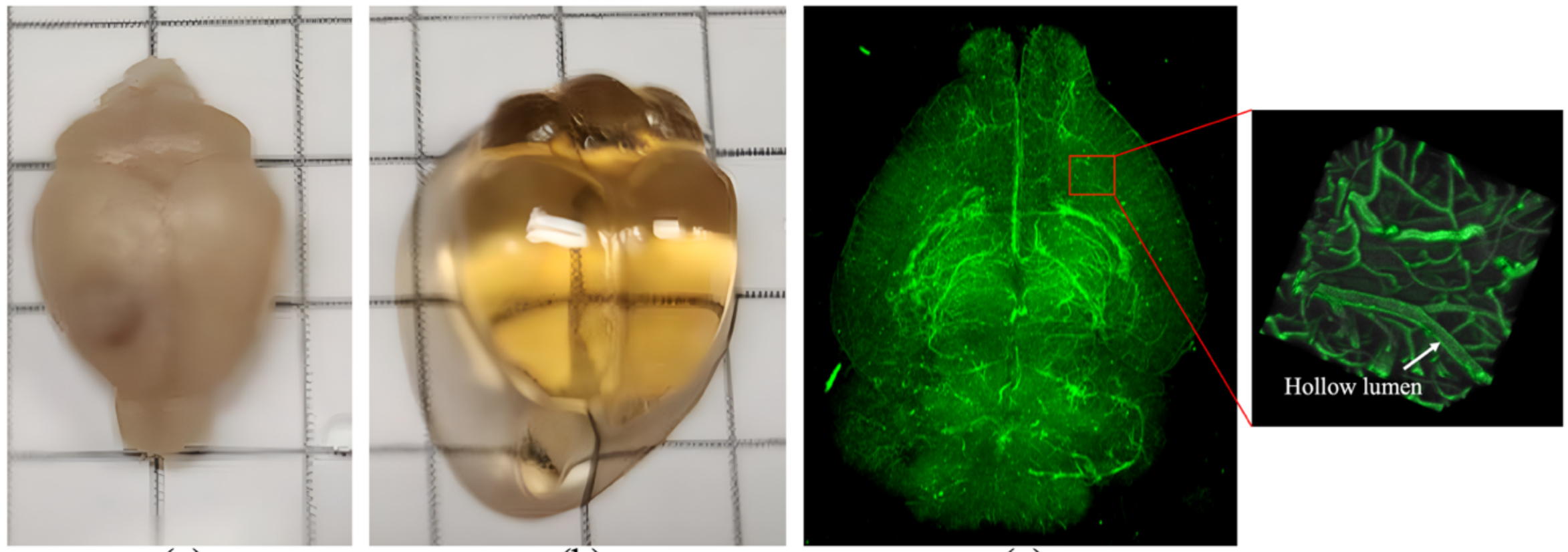

Figure 2. Steps involved in animal preparation and LSFM imaging of a rat brain. (a) A rat brain after skull removal. (b) Same brain after tissue clearing. (c) Three-dimensional (3D) rendition of rat brain vasculature along with a zoomed in volume of interest (VOI) for better representation of the highlighted vessels. This approach enables the visualization of blood vessels of all sizes, including the capillary networks.

In this paper, we present a deep learning (DL) framework that fundamentally redefines MRvF by replacing traditional dictionary matching with a data-driven model trained on experimentally measured vascular networks from rodent whole-brain LSFM images. This approach moves beyond conventional mean vessel size estimation by enabling voxel-wise quantification of intravoxel VSD. LSFM provided whole-brain vascular imaging with sufficient resolution to resolve capillary networks, followed by a custom-developed image processing pipeline for binary segmentation of vascular structures and computation of ground truth CBV and VSD. Voxel-sized VOIs containing segmented blood vessels were then extracted to simulate GESFIDE signals before and after the injection of an intravascular iron-based contrast agent. A fully connected neural network (FCN) was trained to predict quantitative measures of CBV and intravoxel VSD from the ratio of pre- and post-contrast GESFIDE signals. The performance of the DL model was evaluated on both healthy and tumor vasculature by comparing predicted VSD with ground truth measurements, and the accuracy of the mean vessel radius computed from the predicted VSD was assessed against analytically derived VSI (Tropres et al., 2001) measures. The performance of the DL model was evaluated in presence of different level of noise in the simulated GESFIDE signals and compared against the traditional dictionary matching approach. Additionally, the model was validated using a publicly available dataset of segmented mouse brain vasculature (Todorov et al., 2019), further demonstrating its ability to generalize across actual vascular networks.

## 2. METHODOLOGY

In this section, we describe the methods and experimental plans to train and test the DL network for predicting VSD from GESFIDE signal simulated from actual vascular structures. Towards this goal, following materials and methods were used — (1) animal preparation and LSFM imaging, (2) vasculature segmentation and VSD computation, (3) GESFIDE signal simulation, (4) VSD prediction using DL, and (5) experiments and data analysis.

### 2.A Animal Preparation and LSFM Imaging

LSFM images of one healthy rat and mouse, and a rat brain inoculated with a GBM10 patient derived xenograft tumor (Vaubel et al., 2020) were used in this paper. Figure 2 shows the steps involved in animal preparation and LSFM imaging of the rodent brains. First, the animals were sacrificed via trans-cardiac perfusion following a previously published protocol (Scarpelli et al., 2020). Just before perfusion, the blood vessel walls in the brain were highlighted by intravenously administering 100 µL (mouse) or 500 µL (rat) of fluorescently labelled lectin antibody (DyLight 649–labelled Lycopersicon Esculentum Lectin, Vector Laboratories, Burlingame, California). After sacrificing the animal, the brain was removed from the skull (2(a)). Paraformaldehyde-fixed samples then underwent an additional preservation step using SHIELD reagents (LifeCanvas Technologies) as per the manufacturer's instructions (Park et al., 2019). Samples were delipidated using LifeCanvas Technologies Clear+ dilapidation reagents and incubated in 50% EasyIndex (refractive index (RI)=1.52, LifeCanvas Technologies) overnight at 37°C followed by 1 day incubation in 100% EasyIndex for RI matching. The whole-brain vasculature of the rat and mouse were scanned on 3i AxL cleared tissue LSFM scanner (Intelligent Imaging Innovations, Inc., Denver, CO, USA) and SmartSPIM LSFM scanner (Life Canvas Technologies, Cambridge, MA, USA), respectively. Samples imaged on the 3i microscope were imaged in the refractive index matching solution. Samples imaged on the SmartSpim microscope were mounted in 2% ultra-low melt agarose made with EasyIndex, reincubated overnight in EasyIndex, and submerged in EasyIndex matched immersion oil (LifeCanvas Technologies) for imaging (2(b)). The rat brain was acquired at an anisotropic resolution of 1×1×3 µm, whereas the mouse brain was scanned at an isotropic resolution of 1.8 µm. A 3D rendition of the whole rat brain vasculature is shown in (2(c)) and a small VOI is zoomed in for better representation of the highlighted vascular structures. Note that, lectin only stains the vessel wall causing the lumens of larger vessels to appear hollow in LSFM images, whereas smaller vessels appear filled. These hollow lumens result in cavities after segmentation, making accurate segmentation of large vessels more challenging. All animal experiments were conducted after the approval from the Institutional Animal Care and Use Committee (IACUC) at Barrow Neurological Research Institute.

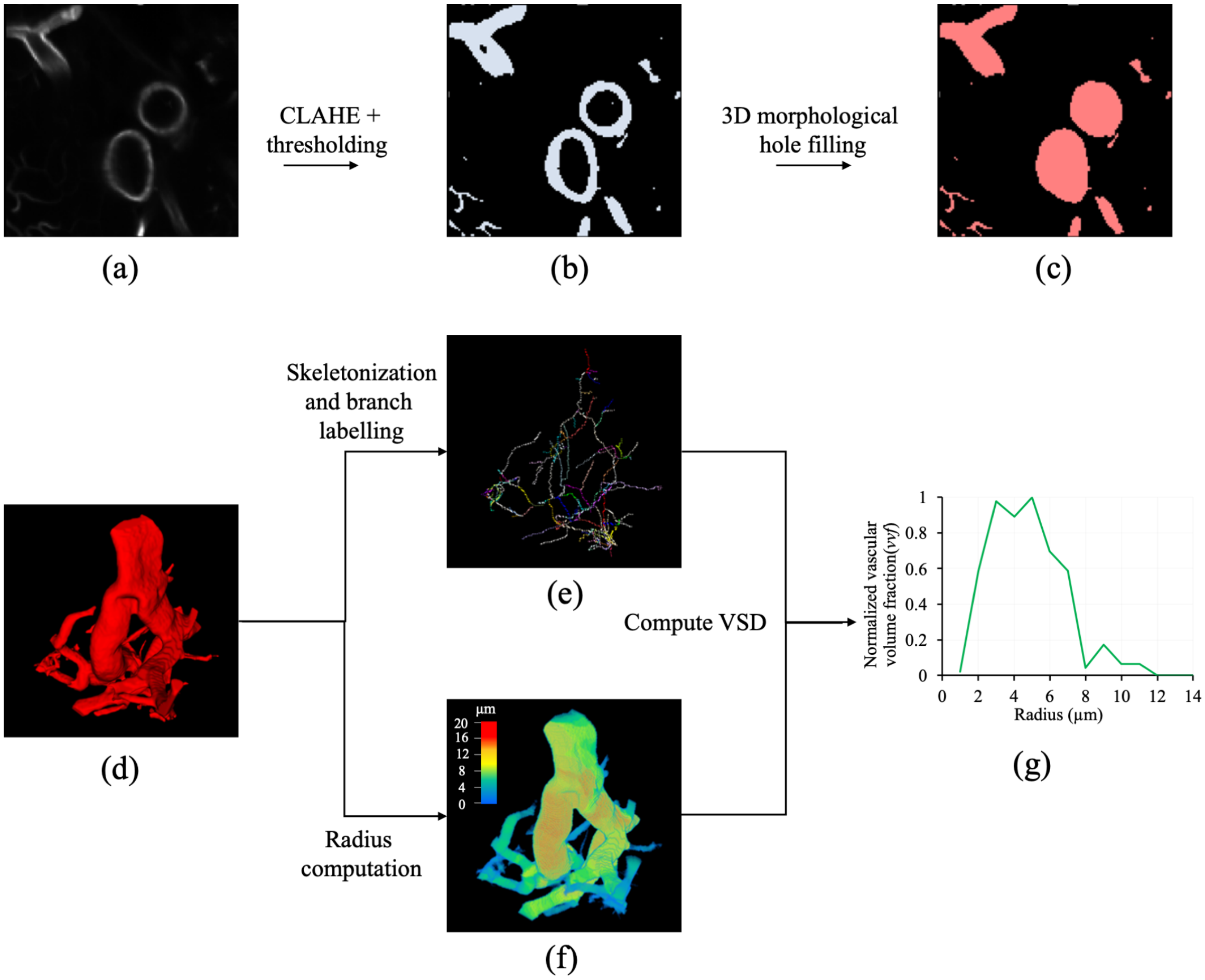

Figure 3. Steps involved in binary segmentation of the vasculature from a LSFM image and computation of the VSD. (a) Axial view of a LSFM VOI with an array size of 123x23x123 (b) Segmented vasculature from (a) after contrast limited adaptive histogram equalization (CLAHE) and binary thresholding (Li & Lee, 1993). (c) Same as (b) but after applying morphological hole filling to close the hollow lumen of the large vessels. (d) 3D rendition of the maximally connected segmented vascular network. (e) Skeleton of (d) where individual vessels are uniquely labelled by distinct colours. (f) Color-coded rendition of voxel-wise radius map of (d). (g) True VSD of (d) computed as the normalized vessel volume fraction (*vvf*) weighted histogram of vessel radius values with a bin size of 1µm . See text for details.

### 2.B Vasculature Segmentation and VSD Computation

LSFM images were first resampled at isotropic resolution of 1.8 µm using linear interpolation. Next, VOIs with an array size of 123 × 123 × 123 were sampled from each LSFM image and passed through an image processing cascade for binary segmentation of the vascular structures and computation of VSD. The image processing cascade was comprised of the following sequential steps — (1) first, contrast limited adaptive histogram equalization (CLAHE) was applied to enhance the contrast of the vascular structures.

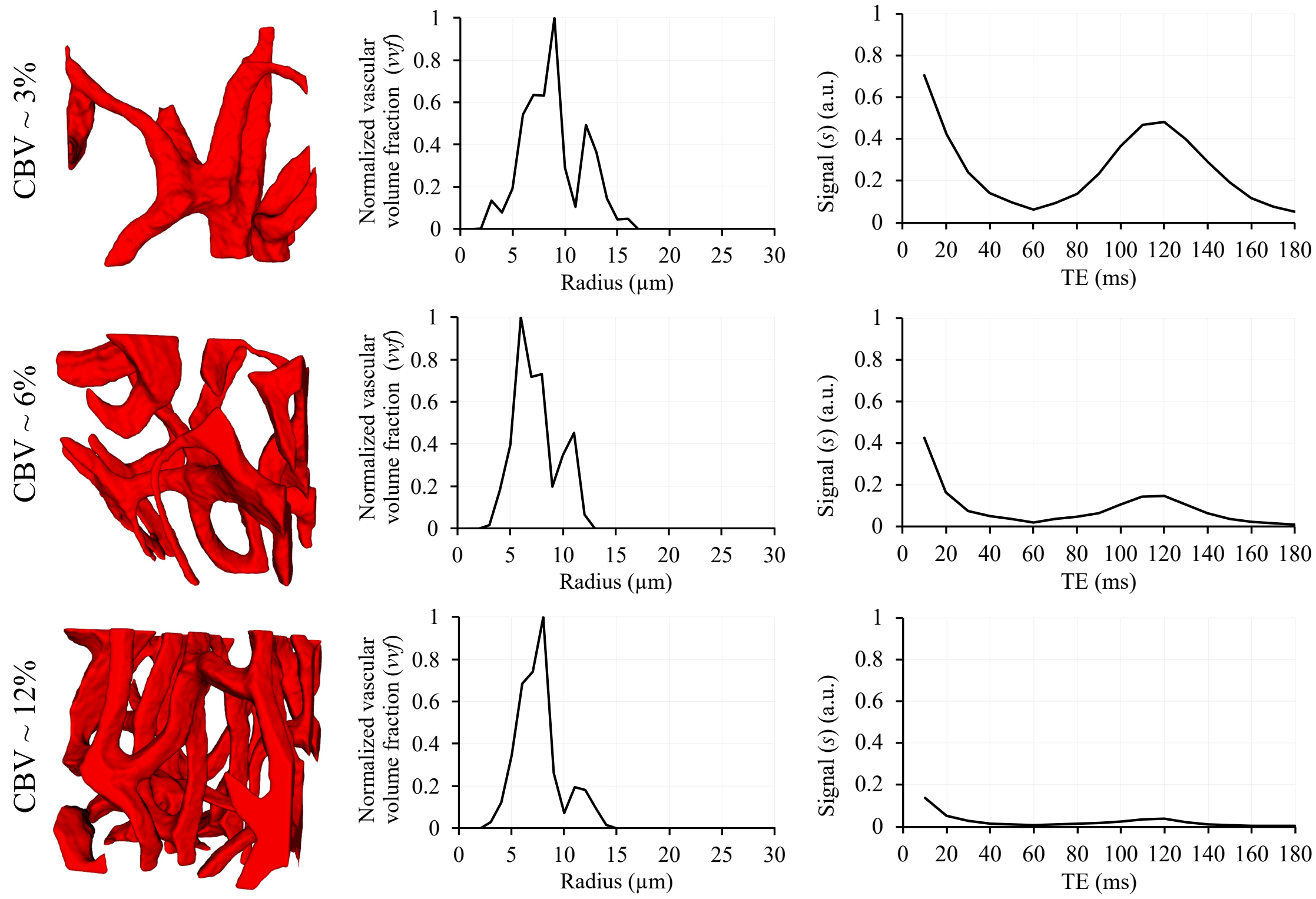

Figure 4. Example VOIs (1st column) extracted from LSFM images with varying CBV and the corresponding VSD (2nd column) and ratio of simulated pre- and post-contrast GESFIDE signals (3rd column). Variations in signal at a specific echo time result from disparities in vessel size, orientation, and CBV.

(2) After contrast enhancement, the vasculature was segmented using a binary thresholding algorithm (Li & Lee, 1993). (3) Next, 3D morphological dilation followed by erosion was applied on the segmented structures to fill the hollow lumen of the segmented large vessels and the maximally connected vascular network was extracted. A spherical kernel with 1µm radius was used for dilation and erosion, and the true CBV was computed as the ratio of non-zero voxels to the total number of voxels in the segmented structure. (4) The skeleton of the segmented vascular structure was extracted using a previously validated algorithm (Lee et al., 1994). Next, each skeletal branch, representing individual vessels, was uniquely labelled using in-house python code. (5) The radius at each skeletal point was computed using a star-line based method (Liu et al., 2014) and the radius of a vessel was determined as the average of the radius values computed at all skeletal points corresponding to that vessel. The mean radius of a VOI was computed as the average radius of all the vessels. (6) As prior studies (Kiselev & Posse, 1999; Kiselev et al., 2005; Yablonskiy & Haacke, 1994) have suggested that the MRI is sensitive to both vessel diameter and shape, we defined the VSD based on the size-differential volume fraction of micro-vessels. Specifically, the *vvf* weighted histogram of the vessel radius values with a bin size of 1µm was computed and normalized by dividing with the maximum bin size to derive the true VSD of a VOI. The *vvf* for a vessel segment is defined as the volume fraction occupied by only that vessel. In other words, instead of counting the number of vessels in a bin, the sum of the *vvf* of each vessel in that bin was computed. So, the height of the $i^{th}$ bin in the VSD represents the normalized sum of the *vvf* ($vvf_i \in [0\ 1]$) corresponding to vessels with radius $i$ µm. Figure 3 illustrates the image processing cascade for vasculature segmentation and VSD computation.

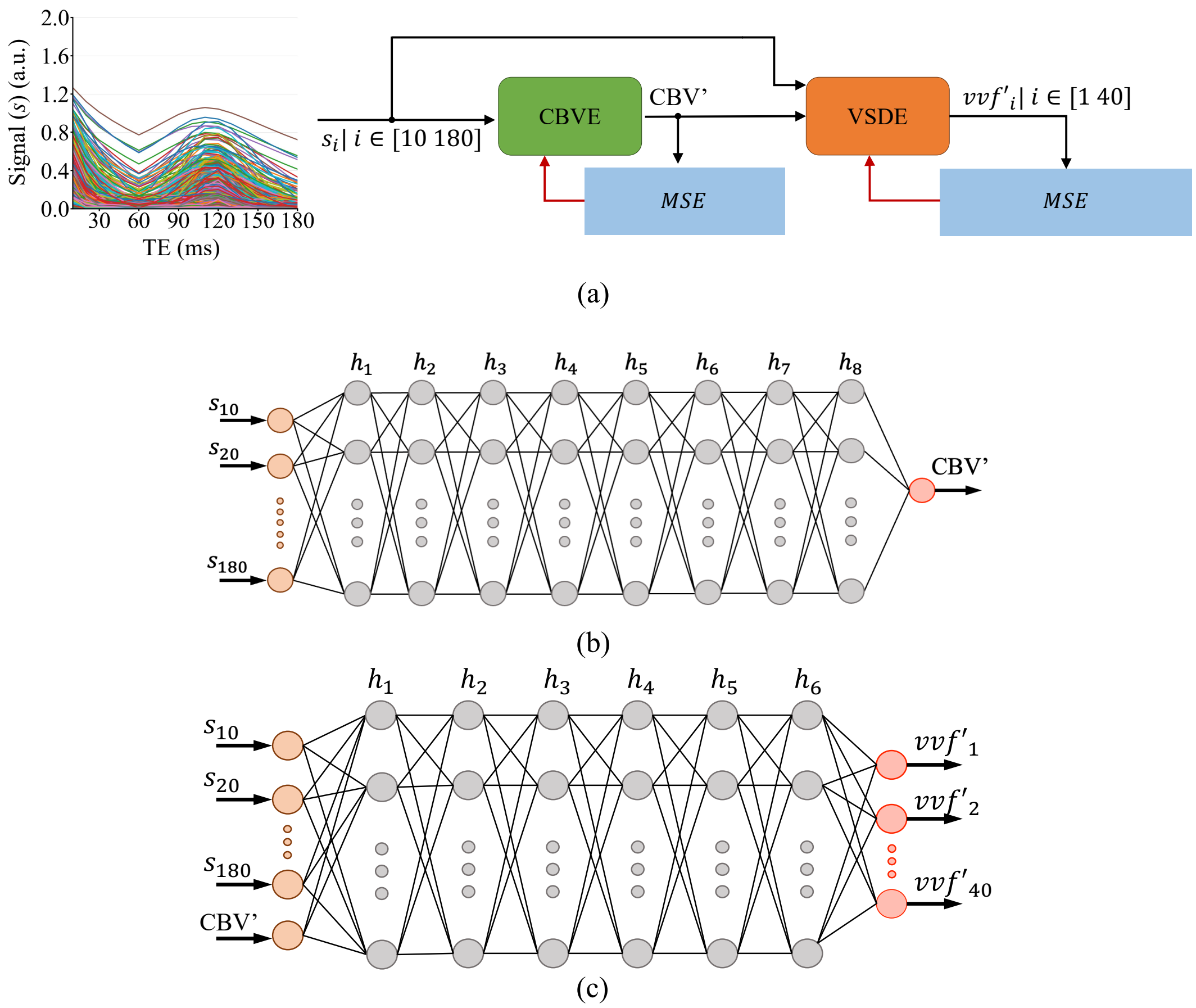

Figure 5. The DL framework used to predict the VSD from a given GESFIDE signal. (a) Two FCNs called the CBVE and VSDE are trained simultaneously to predict the CBV and the VSD from the GESFIDE signal, respectively. (b,c) The network architecture of the CBVE (b) and VSDE (c); see text for details.

**2.C GESFIDE Signal Simulation**

To model the GESFIDE signal evolution, we employed the Finite Perturber Finite Difference Method (FPFDM), a validated computational tool developed by our group, to simulate MR signal changes in realistic 3D tissue structures (Semmineh et al., 2017; Semmineh et al., 2014). The vascular structures derived from LSFM served as the input to the FPFD method, ensuring accurate representation of *in vivo* vascular architecture. VSD imaging is simulated using ferumoxytol, an intravascular iron oxide-based contrast agent. Additional input parameters include static field strength ($B_0$=3 T), susceptibility difference $\Delta\chi$, where $\Delta\chi$ was expressed in the cgs convention (consistent with the 4π-scaled dipole kernel), and a representative $\Delta\chi$=$10^{-6}$ was used for iron-oxide contrast simulation. Using a published ferumoxytol calibration (quantitative susceptibility mapping (QSM) slope ≈ 11.6 ppm·L/g Fe) (Deh et al., 2020), this corresponds to ≈ 0.086 mg Fe/mL (≈ 86 µg/mL; ≈ 1.54 mM Fe). Typical imaging doses of ferumoxytol (∼1–5 mg Fe/kg) produce intravascular concentrations in the same range, given its ∼14–15 h blood-pool half-life. Finally, water diffusion coefficient (D) was set to $10^{-3}$ $mm^2$/s. Using these inputs, a GESFIDE dataset with 18 echo times (10–180 ms) was generated, producing a comprehensive set of signals that serve as input for DL based inference of VSD and CBV. All simulations are conducted using our in-house MATLAB (MathWorks, Natick, MA) code. VSD and simulated GESFIDE signals from VOIs containing vascular structures with varying CBV are shown in Figure 4.

**2.D VSD Prediction using DL**

A two-stage DL network was trained to predict the VSD from the ratio of the simulated pre- and post-contrast GESFIDE signals. The proposed DL network is shown in Figure 5(a). The network is a combination of two FCNs, where the first network, denoted as the CBV estimator (CBVE), predicts the CBV of a VOI from the ratio of the pre- and post-contrast GESFIDE signals and passes it to the second network called the VSD estimator (VSDE). The VSDE takes both the GESFIDE signal and the predicted CBV as input to estimate the VSD. The network is trained in two-steps — first, the CBVE is trained with the objective of minimizing the mean squared error (MSE) between the true and predicted CBV. Next, the weights of the CBVE are set to non-trainable and the VSDE is trained to predict the VSD by minimizing the MSE between the true and predicted VSD.

The CBVE (5(b)) is consisted of 1 input layer, 8 hidden layers ($h_i$), and 1 output layer. The input layer has 18 nodes corresponding to the GESFIDE signal values ($s_i$| $i$=[10 180] ms) at 18 echo times spaced at an interval of 10 ms. The 8 hidden layers consist of 2048, 1024, 512, 256, 128, 64, 16, and 8 nodes, whereas the output layer has only one node corresponding to the predicted CBV (CBV'). The ReLU (Fukushima, 1969) activation function was applied at all hidden layers and the output layer. The VSDE (5(c)) is a combination of 1 input layer, 6 hidden layers, and 1 output layer. The input layer has 19 nodes corresponding to 18 echo times of the GESFIDE signal and the CBV'. The 6 hidden layers have 2048, 1024, 512, 256, 128, and 64 nodes with ReLU activation. The output layer is consisted of 40 nodes where node $i$ outputs the predicted normalized sum of *vvf* ($vvf_i'$) for vessels with radius $i$ µm. A sigmoid (Han & Moraga, 1995) activation function was applied at the output layer to restrict the values of the output nodes between 0 and 1.

16,000 VOIs were sampled from each heathy rat and mouse brain LSFM image, resulting in a total of 32,000 VOIs to train and test the CBVE and VSDE. 680 VOIs with CBV lower than 1% and higher than 40% were removed as they were sampled from the noisy background region and did not include any vascular structures and the, remaining VOIs were split into training, validation, and test dataset in 8:1:1 ratio. Hence, the total number of VOIs in the training, validation, and test set were 25,056, 3,132, and 3,132, respectively. For training, the weights of the two FCNs were initialized following the techniques proposed by He *et al.* (He et al., 2015). Each network was trained using Adam optimizer (Kingma & Ba, 2014) with $\beta_1 = 0.5$ and $\beta_2 = 0.9$ and learning rate of $10^{-4}$ until the training and validation losses converged.

**2.E Experiments and Data Analysis**

The performance of the CBVE and VSDE were first evaluated on the test dataset. The mean $\pm$standard deviation (SD) of the true and predicted CBV were computed and the Pearson correlation (r) between the two sets of CBV values across all the test VOIs (n=3,132) was measured. The mean residual error (MRE) between the true and predicted CBV values was computed and the agreement between the true and predicted CBV values was examined using the Bland-Altman plot.

The accuracy of the predicted VSD was evaluated using Bhattacharyya coefficient (BC) which measures the similarity between the true and predicted VSDs and the mean$\pm$SD of the BC values are reported. For each VOI, the predicted mean vessel radius was computed from the predicted VSD as the *vvf* weighted average of the radius values. The mean$\pm$SD of the true and predicted mean radius as well as the MRE between them were calculated. Furthermore, we have computed VSI from pre- and post-contrast GESFIDE signals ($VSI_{MRI}$) and the LSFM-derived VSD ($VSI_{Histo}$) using the following equations (Tropres et al., 2001; Troprès et al., 2004) —

$$\mathrm{VSI_{MRI}} = 0.425\left(\frac{\mathrm{ADC}}{\gamma\Delta\chi B_0}\right)^{\frac{1}{2}}\left(\frac{\Delta R_2^*}{\Delta R_2}\right)^{\frac{3}{2}} \tag{1}$$

$$\mathrm{VSI_{Histo}} = \left[\int_0^\infty R^{-\frac{2}{3}} f(R)dR\right]^{-\frac{3}{2}}, \tag{2}$$

where ADC is the apparent diffusion coefficient, $\gamma$ is the gyromagnetic constant, $\Delta\chi$ is the susceptibility difference due to the presence of CA, $B_0$ is the external magnetic field, $\Delta R_2^*$ and $\Delta R_2$ are the changes in transverse relaxation rates induced by CA, $f(R)$ is the CBV occupied by vessels of mean radius $R$ such that $\int_0^\infty f(R)dR$=1. Note that, $f(R)$ for a VOI is nothing but the normalized VSD. The analytical equations (Stokes & Quarles, 2016), described in the APPENDIX, were used to compute $\Delta R_2^*$ from the pre- and post-contrast GE signals at the 10 ms and 40 ms echo times. The same equations were applied to compute $\Delta R_2$ from the SE signal at the 120 ms echo time. The mean$\pm$SD of the $VSI_{Histo}$ and $VSI_{MRI}$ values for the test VOIs and the MRE between them are also reported. Also, the agreement between the true mean radius with the predicted mean radius, and $VSI_{Histo}$ and $VSI_{MRI}$ values were evaluated using Bland-Altman plots.

For visual demonstration of the current method, maps of true and predicted CBV were generated, along with BC map for the true and predicted VSDs, at a resolution equivalent to MRI, across a stack of 123 axial slices of segmented mouse brain vasculature. The maps of true and predicted mean radius and $VSI_{MRI}$ values over the same stack were also computed. To achieve this, the stack was divided into non-overlapping VOIs, each with an array size of 123×123×123. The true parameter values were computed using LSFM-based algorithms, while the trained model was used to predict the corresponding parameters for each VOI. The computed values were then assigned back to all the voxels within the corresponding VOIs in the binary LSFM vasculature stack, preserving spatial resolution. Finally, parameter maps for the entire stack were assembled by stitching together the corresponding VOI-based maps.

We compared the performance of the DL models in predicting CBV and VSD against the traditional dictionary-matching approach (Christen et al., 2014). Specifically, each signal in the test dataset was compared to the signals in the training dataset by computing the coefficients of determination $\mathcal{R}^2$ and the CBV and VSD corresponding to the signal producing highest $\mathcal{R}^2$ was selected. The mean ± SD of the predicted CBV, VSD, and mean radius were reported. Pearson correlation (r) and MRE between true and predicted CBV and mean radius were calculated, along with the mean ± SD of BC values for VSDs.

To further assess the feasibility of applying the trained DL models to *in vivo* data, we evaluated its performance on GESFIDE signals with varying levels of signal-to-noise ratio (SNR). Specifically, we added white Gaussian noise to the test dataset signals at SNR levels of 15, 30, 45, and 60 dB, and applied the trained CBVE and VSDE models to predict CBV and VSD, respectively. For each SNR, the mean±SD of the predicted CBV and mean vessel radius were computed. We also computed the Pearson correlation (r) and MRE between true and predicted CBV. In addition, we computed the mean±SD of BC values between the true and predicted VSDs, as well as the MRE between the true and predicted mean vessel radius.

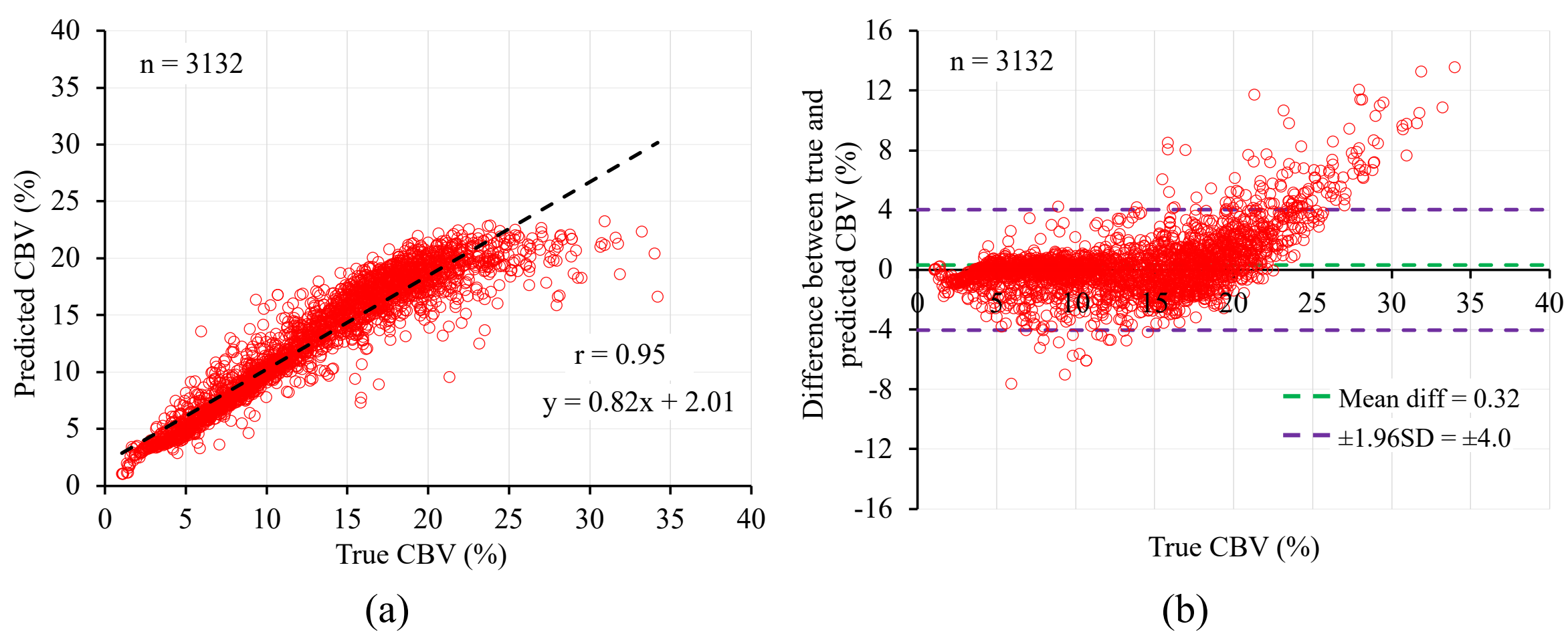

Figure 6. The Pearson correlation (a) and Bland-Altman plots (b) of true and predicted CBV values (n=3,132). The predicted CBV values were not only highly correlated (r=0.95) but also very similar (mean difference of 0.32%) to the true CBV values.

The DL models were tested on a publicly available dataset (Todorov et al., 2019) containing VOIs of segmented vasculature from a mouse brain to examine the generalizability of the DL models. 1,000 VOIs were randomly selected and cropped to the array size of $123 \times 123 \times 123$ at 3 µm isotropic resolution. Additionally, to examine the performance of the DL models on tumor vasculature, we sampled 706 VOIs of array size 123×123×123 from the tumor and peritumoral regions (supplementary Figure S1) of the LSFM image (1.8 µm isotropic) of a rat brain inoculated with a GBM10 patient derived xenograft tumor and segmented the vasculature using the segmentation algorithm described in Section 2.B. In both cases, the simulated GESFIDE signal from each VOI was passed through the trained CBVE and VSDE models to predict the CBV and VSD, respectively. The mean±SD of the true and predicted CBV and mean vessel radius were computed. The Pearson correlation (r) and MRE between true and predicted CBV were computed. The mean±SD of BC values between the true and predicted VSDs was computed and the MRE between the true and predicted mean vessel radius were measured. The mean±SD of the $VSI_{Histo}$ and $VSI_{MRI}$ values for the VOIs and the MRE between them are also reported.

## 3. RESULTS

### 3.A Evaluation of CBV Prediction Performance on Test Set

The mean±SD of the true and predicted CBV values for the test VOIs was 13.2±6.2 and 12.9±5.4%, respectively with a MRE of 10%. The scatter plot of true and predicted CBV values (n=3,132) are shown in Figure 6(a). A strong linear correlation (r = 0.95) can be observed between the two sets of values with the trend-line having a slope and intercept of 0.82 and 2.01, respectively. The Bland-Altman plot of the difference between true and predicted CBV values are shown in 6(b). The mean difference between the true and predicted CBV was 0.32% and 97% of the residuals were within±1.96SD, i.e. ±4.0%, of the mean difference. Figure 7 shows the color-coded true (7(a)) and predicted (7(b)) CBV maps over an entire axial slice of the mouse brain. Visual similarity between the true and predicted CBV maps can be observed which is further supported by the difference image in 7(c), where most of the pixels demonstrate a very low difference between the true and predicted CBV values.

### 3.B Evaluation of VSD Prediction Performance on Test set

The qualitative results of VSD prediction for 12 VOIs with CBV varying from 2.2 to 24% are shown in Figure 8. Significant overlap between the true (green) and predicted (red) VSDs can be observed for all 12 VOIs. The BC values for the test VOIs (n=3,132) are plotted against their true mean radius in Figure 9(a). The mean±SD of the BC values were 0.87±0.09 and 60% of the VOIs showed a BC of $\geq$ 0.87 while the BC for 72% of the VOIs were within ±1SD of the mean value. The mean±SD of the true and predicted mean radius were 7.3±2.5 and 7.4±1.0 µm, respectively with 28% MRE between them. The Bland-Altman plot of the true and predicted mean radius (9(b)) shows a mean difference of -0.16 µm with 95% of the residuals falling within ±1.96SD, i.e. ±5.3 µm, of the mean difference. The $VSI_{Histo}$ and $VSI_{MRI}$ for each VOI was computed using the Equations (1 & 2) where the values of $ADC$, $\gamma$, $B_0$, and $\Delta\chi$ were set to 1

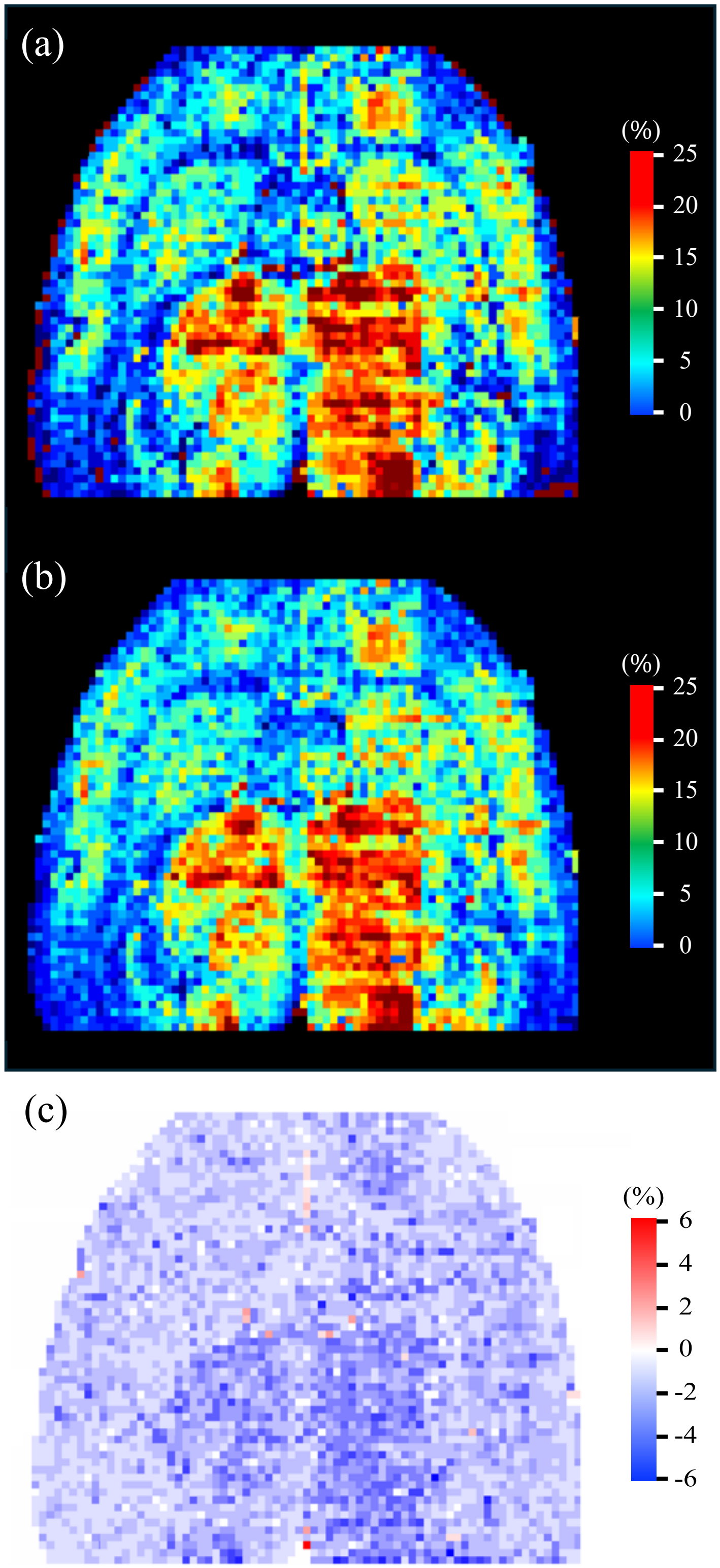

Figure 7. Color-coded maps of true (a) and predicted (b) CBV values over an entire axial slice of mouse whole brain LSFM image. The difference map of the true and predicted CBV values is shown in (c).

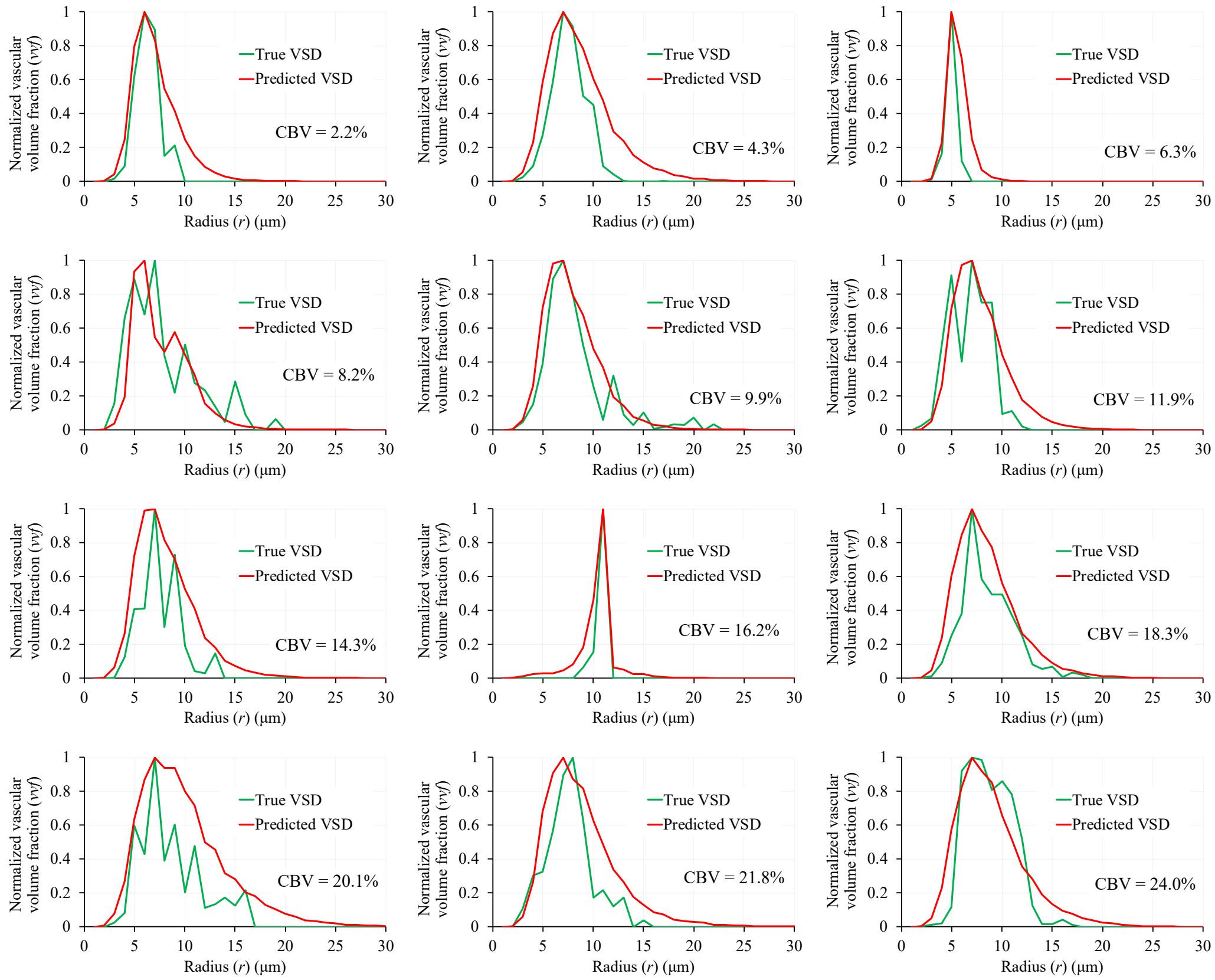

Figure 8. The true (green) and predicted (red) VSDs for 12 VOIs with CBV varying from 2.2 to 24%. Significant overlap between the true and predicted VSDs are noticeable despite the true VSD being noisy and of varying shape.

$\mu m^2$/ms, $4.258 \times 10^{-7}$ $s^{-1}T^{-1}$, 3 T, and $10^{-6}$, respectively. The mean±SD of the $VSI_{Histo}$ and $VSI_{MRI}$ were 6.7±2.2 μm and 8.1±3.9 μm, respectively with a 55% MRE between them. The Bland-Altman plot of

(9(c)) shows a mean difference of -1.4μm between the true mean radius and VSI with 95% of the residuals falling within ±1.96SD, i.e. ±8.4 μm, of the mean difference.

The color-coded BC map derived from the true and predicted VSDs over an entire axial slice of mouse brain is shown in Figure 10(a). The BC map has mostly red pixels (BC > 0.80) that further demonstrates high similarity between the true and predicted VSDs. The mean BC over the entire axial slice was 0.78. The color-coded maps of true (10(b)) and predicted mean radius (10(c)), and $VSI_{MRI}$ (10(d)) are also shown. Visually, the predicted mean radius appears to be highly similar to the true mean radius compared to the $VSI_{MRI}$ values across the entire axial slice. Visual comparison of the difference map between the true and predicted mean radius (10(e)), and true mean radius and $VSI_{MRI}$ (10(f)) further demonstrates that the predicted radius is closer to the true value than the $VSI_{MRI}$.

### 3.C Comparison with Traditional Dictionary-matching Approach

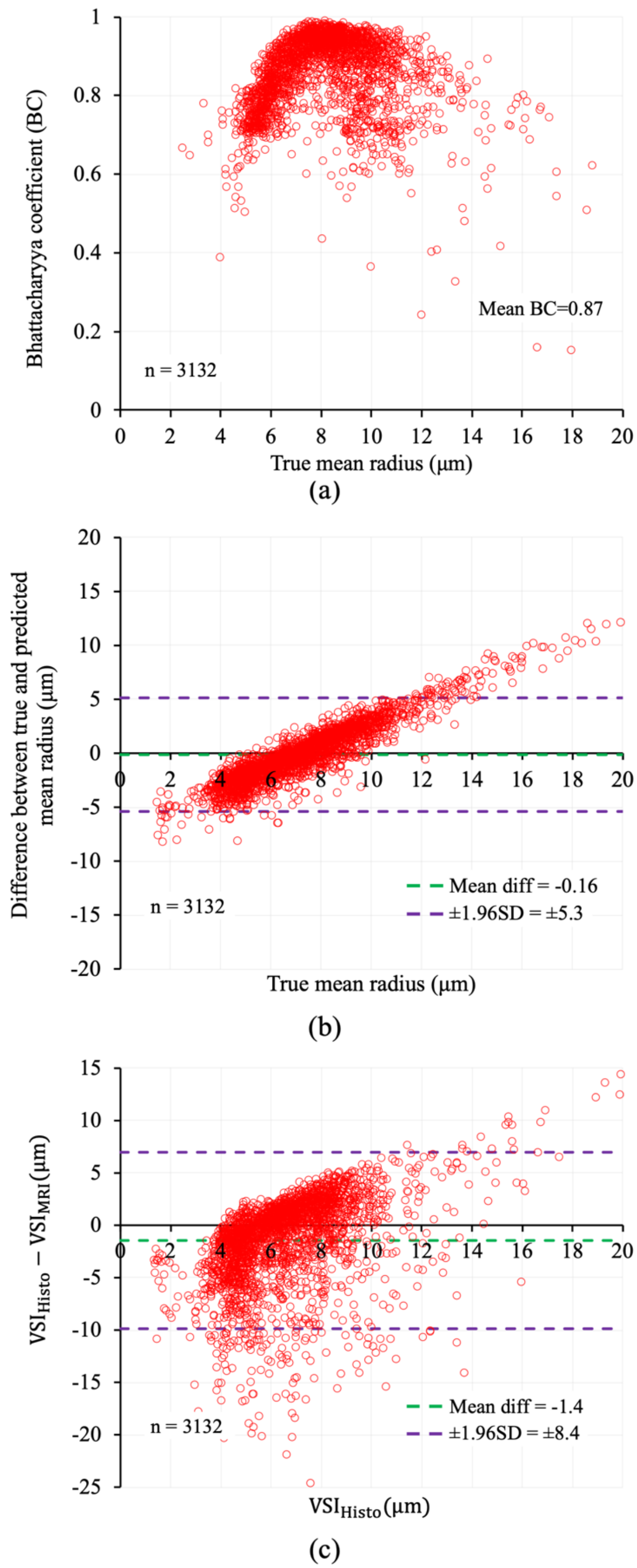

Figure 9. The results of quantitative evaluation for the VSDE. (a) Distribution of the BC values against the true mean radius for the test VOIs (n=3,132). (b) The Bland-Altman plot of difference between true and predicted mean radius. (c) Same as (b) but for the difference between $VSI_{Histo}$ and $VSI_{MRI}$.

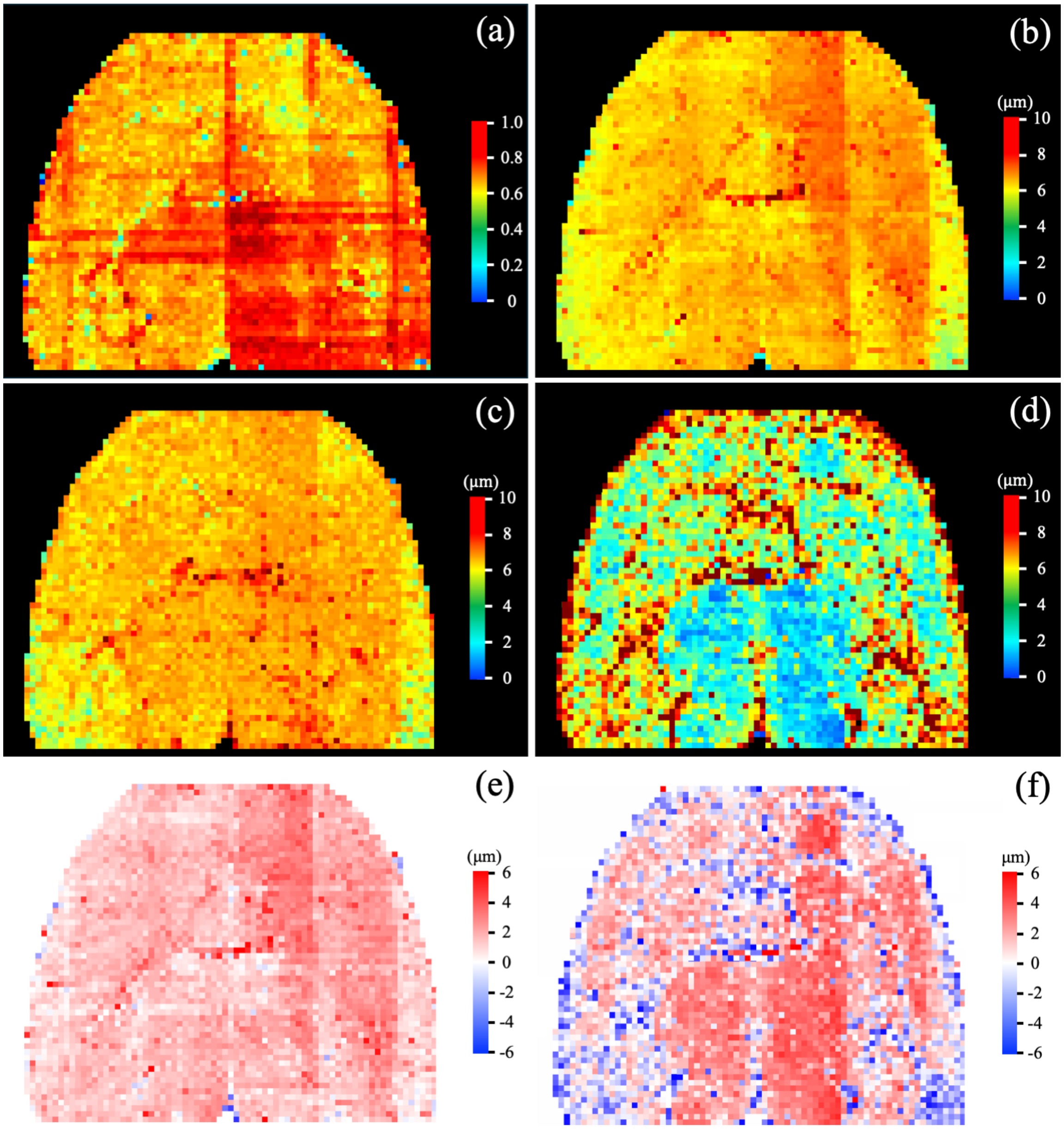

Figure 10. Qualitative results of BC, mean vessel radius, and $VSI_{MRI}$ computation on an entire axial slice of mouse whole brain LSFM image. (a-d) Color-coded maps of the BC (a), true (b) and predicted (c) mean radius, and $VSI_{MRI}$ (d) values. (e,f) The residual map of true and predicted mean radius (e) and the true mean radius and $VSI_{MRI}$ (f).

The DL models outperformed the traditional dictionary matching approach in predicting both CBV and VSD on the test dataset (n=3,132). The mean±SD of the CBV values predicted using the dictionary matching approach was 12.8±7.3%. The predicted CBV values showed a moderate linear correlation (r=0.69) and MRE of 21% with the true CBV values. The mean±SD of the BC values were 0.17±0.02. The mean±SD of the predicted mean radius were 1.2±0.62 μm and showed 84% MRE against the true mean radius values.

### 3.D Influence of Noise on the Performance of the DL Models

The mean±SD of the CBVE predicted CBV values at SNR 15, 30, 45, and 60 were 14.6±5.6, 14.3±5.7, 13.5±6.1, and 13.3±6.1%, respectively. The linear correlation between the true and predicted CBV values at 4 SNR levels were 0.83, 0.85,0.89, and 0.92, respectively with MRE being 24, 17, 15, and 12%, respectively. The mean±SD of the BC values for the 4 SNR levels were 0.82±0.13, 0.83±0.12, 0.84±0.12, and 0.85±0.13, respectively. The mean±SD of the predicted mean radius for SNR 15, 30, 45, and 60 were 7.9±1.5, 7.7±1.3, 7.6±1.3, and 7.6±1.3 μm, respectively, whereas the MRE against the true mean radius values were 30, 29, 29, and 28.5%, respectively.

### 3.E Performance of the DL Models on a Publicly Available Dataset

The mean±SD of the true and predicted CBV values (n=1,000) for the publicly available dataset were 6.6±4.3 and 7.7±4.7%, respectively with a MRE of 21%. The linear correlation of 0.96 was observed between the true and predicted CBV. The trend-line closely followed the identity line with a slope and intercept of 1.0 and 0.86, respectively. The mean±SD of the BC values were 0.84±0.10 and 59% of the VOIs showed a BC of $\geq$ 0.84 while the BC for 89% of the VOIs were within ±1SD of the mean value. The mean±SD of the true and predicted mean radius were 7.1±2.5 and 7.7±0.82 μm, respectively with 26% MRE between them. The mean±SD of $VSI_{Histo}$ and $VSI_{MRI}$ were 5.3±1.9 μm and 11.4±5.4 μm, respectively. The MRE between the $VSI_{Histo}$ and $VSI_{MRI}$ was 123%.

### 3.F Performance of the DL Models on Tumor Brain

The mean±SD of the true and predicted CBV values (n=706) for the tumor VOIs were 15.2±4.5 and 15.1±4.2%, respectively with a MRE of 17%. The linear correlation of 0.78 was observed between the true and predicted CBV. The correlation and Bland-Altman plots for the true and predicted CBV values are presented in supplemental Figure S2. The mean±SD of the BC values were 0.82±0.12 and 60% of the VOIs showed a BC of $\geq$ 0.82 while the BC for 78% of the VOIs were within ±1SD of the mean value. The mean±SD of the true and predicted mean radius were 7.3±1.6 and 9.6±1.8 μm, respectively with 43% MRE between them. The mean±SD of the $VSI_{Histo}$ and $VSI_{MRI}$ were 5.7±1.5 and 5.1±3.6 μm, respectively, while the MRE between them was 47%. The Bland-Altman plots for true and predicted mean radius values as well as the $VSI_{Histo}$ and $VSI_{MRI}$ values are presented in supplemental Figure S3.

## 4. DISCUSSION

The results of this study provide strong evidence that imaging intravoxel VSD is feasible by training a DL network using pre- and post-contrast GESFIDE MRI data simulated from actual vascular structure extracted from rodent brain. A significant feature of the approach is that it intrinsically decouples the complex relationship between voxel-wise, CA concentration, heterogeneous vascular architecture, and the measured changes in transverse relaxation rates. With traditional DSC-MRI, this relationship is unknown, varies across voxels, and prevents absolute quantitation of the derived hemodynamic parameters. Practically, the *in vivo* data needed to image VSD is equivalent to MRvF; steady-state GESFIDE data collected prior to and after the injection of an iron-oxide based intravascular CA.

Several histopathological studies (Chakhoyan et al., 2019; Jain et al., 2011; Pathak et al., 2001) have reported weak to moderate linear correlations ($r \in [0.42\ 0.74]$) between the MRI-derived relative CBV measure and histology-derived fractional CBV and vessel density. In contrast, the high linear correlation (r=0.95) and low MRE (10%) between the true and predicted CBV observed in this study demonstrate that DL models trained on the ratio of pre-and post-contrast GESFIDE signal simulated from realistic 3D vascular structures can accurately recover true CBV, rather than a relative surrogate, unlike conventional DSC-MRI rCBV measures (Boxerman et al., 2020). Previous MRvF studies (Boux et al., 2021; Christen et al., 2014) evaluated CBV accuracy using virtual VOIs containing 2D cylindrical vessels of uniform radius and reported relatively lower MREs (~4%) but also noted increasing error at higher CBV values. The Bland–Altman plot in Figure 6(b) indicates minimal mean difference across the full CBV range, demonstrating an absence of systematic bias toward higher CBV values. Note that the elevated true CBV values (>10%) reported herein arise from the small LSFM-derived simulation VOIs (123 × 123 × 123 voxels at 1.8 µm resolution, i.e., ≈221 µm isotropic). At this scale, individual VOIs may be dominated by one or a few vessels, particularly those sampled near larger branches resulting in locally high CBV. This VOI size was selected to preserve microvascular detail while keeping computation time manageable. When LSFM voxels are aggregated to the scale of an MRI voxel (≈1 mm isotropic), the CBV distribution contracts to physiologic levels (~3–5%), consistent with the previously reported CBV values (Cassot et al., 2006).

High BC values between the true and predicted VSDs demonstrate that the DL model is sensitive to the subtle variations in GESFIDE signal caused by the underlying vascular structure in a VOI and accurately predicts VSD from the signal. The mean vessel radius computed from the predicted VSD was closer to the true mean radius than the $VSI_{MRI}$ values. However, the MRE between the true and predicted mean radius (28%) was higher than the 9% error reported in the first MRvF study (Christen et al., 2014), likely reflecting the increased difficulty of differentiating GESFIDE signals generated from realistic 3D vascular geometries with non-uniform radii, branching, tortuosity, and orientation heterogeneity compared with simplistic 2D cylindrical structures. Notably, more recent MRvF implementations that incorporate realistic microvascular networks derived from angiograms or whole-brain microscopy/LSFM data have primarily evaluated their performance in terms of physiological plausibility of parameter maps and agreement with literature values or independent measurements, rather than reporting voxel-wise percentage errors relative to a microvascular ground-truth (Barrier et al., 2024; Delphin et al., 2024). In this context, our direct comparison of predicted mean radius against LSFM-derived radius distributions and the associated 28% MRE represents a stricter quantitative benchmark than is typically reported for LSFM-based MRvF frameworks. The Bland-Altman plots in Figure 9 show that the $VSI_{MRI}$ under- or over-estimates $VSI_{Histo}$ and the variability of the residuals between $VSI_{Histo}$ and $VSI_{MRI}$ are ~2 times higher than the variability of the residuals between true and predicted mean radius. This result is consistent with prior MRI–histology comparisons (Kellner et al., 2015; Troprès et al., 2015), which also report under- or over-estimation of vessel sizes by MRI-derived VSI.

The lower agreement between the MRI-derived $VSI_{MRI}$ and the LSFM-based $VSI_{Histo}$ values arise from the simplifying assumptions behind Equation (1). For comparison, the $VSI_{Histo}$ was computed directly from the size-differential vessel volume fraction using Equation (2). The analytical VSI model assumes infinitely long cylinders of uniform radius, random orientation, the validity of the static-dephasing regime for $\Delta R_2^*$ (Yablonskiy & Haacke, 1994), and the slow-diffusion approximation for $\Delta R_2$ (Kiselev & Posse, 1999). These conditions are only approximately satisfied at mesoscopic MRI voxel scales. Prior MRI-histology studies have accordingly reported moderate correlations (typically $r \in [0.5–0.7]$) at the VOI level (Chakhoyan et al., 2019; Kellner et al., 2015; Lemasson et al., 2013), where each voxel contains many vessels and the effective-medium assumptions hold. In contrast, the LSFM-derived test VOIs represent substantially smaller vascular neighbourhoods (~221 $\mu m^3$ isotropic) with pronounced heterogeneity in vessel radius, branching geometry, and CBV, conditions under which the VSI analytical model is less valid, and its dynamic range becomes compressed. Consistent with this interpretation, VSI correlations improved ($r \approx 0.5$) when applied to a larger-voxel external LSFM dataset (~369 $\mu m^3$ effective VOI size). These observations agree with recent Monte-Carlo–based *in silico* and *in vivo* analyses demonstrating that VSI accuracy and sensitivity degrade markedly below ~250 μm spatial resolution due to partial-volume averaging and heterogeneous vessel distributions (Lee et al., 2025). Thus, the weak agreement between $VSI_{Histo}$ and $VSI_{MRI}$ observed in the test dataset reflects a breakdown of the analytical VSI assumptions rather than errors in LSFM processing or signal simulation. In contrast, the DL-predicted VSD does not depend on these geometric or diffusion approximations and therefore yields more reliable radius estimation across diverse vascular architectures.

Both CBVE and VSDE demonstrated superior performance relative to the traditional dictionary-matching approach in predicting both CBV and VSD. However, the dictionary-based method exhibited relatively greater accuracy in the estimation of CBV than VSD. This finding may be explained by the more linear relationship between the simulated GESFIDE signal and CBV, compared with the subtler and more complex relationship between the signal and VSD. For example, in Figure 4, the magnitude of the simulated GESFIDE signal decreases as CBV increases, whereas no comparable linear trend can be identified between the signal values and VSD. Hence, the substantially elevated MRE observed between the true and predicted mean vessel radius can be attributed to the inherent limitations of the dictionary-matching approach in reliably capturing VSD distributions.

The DL models demonstrated robust performance in predicting both CBV and VSD even in the presence of substantial noise within the GESFIDE signals, with prediction accuracy improving as the SNR increased. At an SNR of 15, the linear correlation between true and predicted CBV decreased by 12.6% relative to the correlation measured in the absence of noise. More notably, the reduction in mean BC between the true and predicted VSD distributions at SNR 15 was limited to 5.7% compared with the noiseless condition. These findings underscore the feasibility of applying the proposed DL models for estimating CBV and VSD from noisy *in vivo* GESFIDE data. Nonetheless, to achieve optimal performance, the models must be retrained using simulated noisy GESFIDE signals in conjunction with ground-truth CBV and VSD values derived from LSFM images.

The quantitative evaluative results on the publicly available dataset were comparable to those observed in the test dataset. These observations demonstrate that the GESFIDE simulation and VSD computation algorithms, and trained DL models are not biased towards the tissue clearing process, LSFM imaging parameters, and vessel segmentation algorithm, rendering them readily applicable across different datasets without retraining the DL models. Although the DL models demonstrated reduced performance on tumor VOIs relative to healthy VOIs, the lower linear correlations and mean BC values, together with the higher MREs between the true and predicted CBV and VSDs, are likely attributable to the abnormal vascular morphology in tumor regions. These atypical vessel characteristics produce MR signals that were not represented in the DL models trained exclusively on healthy VOIs. Therefore, to achieve optimal performance, the DL models should be trained on a large number of tumor VOIs in addition to healthy VOIs. However, this was not feasible in the present study due to the limited availability of LSFM images of tumor brains (only one tumor brain).

There are few limitations of the current study that should be clarified. First, due to noise and the resolution limitation of the LSFM image, very small (<1.8 μm) capillaries may merge and result in erroneous computation of skeletal points and vessel radius. As the computation of the true VSD is sensitive to both the localization error in the skeletal points and the over or under segmentation of the vascular structures, the DL model will have intrinsic learning error that can only be fixed by using more accurately segmented vascular structure. However, developing a highly accurate vessel segmentation algorithm is beyond the scope of the current paper. Second, the maximum vessel radius observed in 32,000 VOIs used for training and validation of the DL model was 20 μm. So, the model needs to be validated on vascular structures with radius higher than 20 μm, specifically for translating the method into human brain where the vessel radius reaches up to 3 mm. Third, the significant heterogeneity in vascular morphology such as vessel radius, length, and density among different organs may require the development of organ specific DL models for accurate VSD prediction. Fourth, we simulate steady-state intravascular susceptibility and feed the entire multi-echo GESFIDE manifold (multiple GEs, ASEs, and SEs) into an end-to-end network that learns VSD without prescribing how GE or SE depend on $\chi$. In this setup, the assumed $\Delta\chi$ mainly sets a global contrast scale. We did not systematically vary contrast concentration or acquisition timings (e.g., echo spacing, ASE offsets) to test their impact on radius scale or throughput; a focused sensitivity study is a reasonable next step but lies outside the scope of this work. Fifth, $\Delta B_0$ was computed from the intravoxel vasculature, and diffusion was solved with a finite-difference method using periodic (cyclic) boundary conditions to mitigate edge effects and tile the unit cell for spin motion (Semmineh et al., 2014; Xu et al., 2007). Cyclic boundaries together with a unit-cell scale chosen to capture long-range effects are widely used and yield stable GE/SE predictions, with relaxation dominated by intravoxel structure (Berman et al., 2025). Within the unit cell, overlapping fields from many vessels are fully accounted for via finite-perturber finite-difference method (Semmineh et al., 2014). We acknowledge that explicit neighboring geometry is not included in $\Delta B_0$ here; a future sensitivity analysis will expand the field domain to assess any residual far-field influence. Sixth, in our simulation, diffusion is modelled by solving the Bloch–Torrey equation (finite differences with periodic boundaries), but diffusivities were held fixed to isolate vascular-size contrast; we note a planned diffusivity/permeability sweep as future work. Under steady-state conditions, susceptibility in monocrystalline iron oxide nanoparticles is primarily governed by super Para magnetism, and using the standard relation $\Delta\chi_{\mathrm{blood}} \approx \mathrm{Hct}\Delta\chi_0(1 - \mathrm{SO_2})$, physiological oxygenation changes contribute ~10–20× less susceptibility than typical MION levels; accordingly, we fixed $\mathrm{SO_2}$ and focused on VSD and CBV. As a future work we will extend the simulation library to sample $\mathrm{SO_2}$ (or add oxygenation-sensitive echoes) in a targeted sensitivity analysis.

In summary, this is the first study to develop and validate a prediction model for estimating VSD from pre- and post-contrast GESFIDE signals simulated from 3D vascular structures extracted from the LSFM images of whole rodent brains. Although extensive *in vivo* validation of the DL model is required, the findings of the *ex vivo* experiment presented in this paper have shown the potential of VSD imaging as a new imaging approach to quantitatively characterize vascular remodelling associated with disease and therapy. Non-invasive VSD imaging may enable novel insights into healthy and disease tissue vasculature microstructure and heterogeneity without the risks of invasive biopsies. With the goal of incorporating VSD into clinical workflows for brain tumor patients, future studies will explore whether the developed algorithms can be applied, at least to some degree, to DSC-MRI scans acquired with clinical Gadolinium-based contrast agents and spin and gradient echo (SAGE) type pulse sequences (Keeling et al., 2024; Skinner et al., 2014). The SAGE pulse sequence is an echo planar version of GESFIDE and provides a reduced number of echo times. Once matured, VSD imaging could improve diagnostic accuracy, guide personalized therapies and serve as biomarkers of therapeutic response. Importantly, these innovative imaging techniques have potential applications across most organs, diseases, and mammalian species.

## 5. AUTHOR CONTRIBUTIONS

N.B.S., I.G. 'Conceptualized the project, developed the MR signal simulation tool, image analysis algorithms, and deep learning model, designed the experiments, analyzed the data, and wrote the manuscript.', D.H. and A.C. 'Acquired the LSFM images and wrote the manuscript', J.L.B. 'Conceptualized the project, analyzed the data, and wrote/edited the manuscript.', C.C.Q. 'Conceptualized, supervised, acquired funds for the project, and developed the methodology, designed experiments, analyzed data, and wrote/edited the manuscript'. All the authors discussed the results and commented on the manuscript. All authors have approved the final version of the manuscript.

## 6. DECLARATION OF COMPETING INTEREST

The authors declare no competing financial interests.

## 7. ACKNOWLEDGEMENTS

This work was supported by Cancer Prevention and Research Institute of Texas (CPRIT) RR220038 (C. Chad Quarles is a CPRIT scholar in cancer research), MD Anderson's Advanced Technology Genomics Core (supported by NIH 1S10OD024977-01 and NCI P30CA0166722), Research Histology Core Laboratory (supported by NCI P30CA0166722), Advanced Microscopy Core (supported by NIH S10RR029552), and Advanced Cytometry & Sorting Facility (supported by NCI P30CA0166722).

## 8. DATA AVAILABILITY

While the full dataset cannot be shared due to its large size (~1TB), sample data will be provided. The complete raw and segmented LSFM images will be available upon reasonable request.

## 9. CODE AVAILABILITY

The source code and sample test data used in this paper is available online at https://github.com/iguha94/VSD_Pipeline.git.

## APPENDIX

The analytical equations to estimate T1 insensitive $\Delta R_2^*$ and $\Delta R_2$ using gradient echo (GE) at timepoints $TE_1$ and $TE_2$, and spin echo (SE) at timepoints $T_{\text{SE}}$ are described here. The $\Delta R_2^*$ is computed using the following equation:

$$\Delta R_2^*(t) = \frac{1}{TE_2 - TE_1} \left( \ln\left(\frac{S_{TE_{2,pre}}}{S_{TE_2}}\right) - \ln\left(\frac{S_{TE_{1,pre}}}{S_{TE_1}}\right)\right), \tag{1}$$

Where, $S_{TE_{1,pre}}$ and $S_{TE_{2,pre}}$ are the baseline GE signals at echo time $TE_1$ and $TE_2$, respectively. $S_{TE_1}$ and $S_{TE_2}$ are the GE signals at echo time $TE_1$ and $TE_2$, after contrast-agent (CA) arrival.

Next, the T1-weighted signal extrapolated to TE=0 is computed using the following equation:

$$S_{TE=0} = S_{TE_1} . \left(\frac{S_{TE_1}}{S_{TE_2}}\right)^{\frac{TE_1}{TE_2 - TE_1}} \tag{2}$$

Finally, $\Delta R_2$ is computed using $S_{TE=0}$ and the following equation

$$\Delta R_2(t) = \frac{1}{TE_{SE}} \left( \ln\left(\frac{S_{TE_{SE,pre}}}{S_{TE_{SE}}}\right) - \ln\left(\frac{S_{TE=0,pre}}{S_{TE=0}}\right)\right), \tag{3}$$

Where, $S_{TE_{SE,pre}}$ and $S_{TE_{SE}}$ are the pre- and post-contrast SE signals, and $S_{TE=0,pre}$ and $S_{TE=0}$ are the pre- and post-contrast T1-weighted signal extrapolated to TE=0.